\documentclass[]{aa}
\usepackage{amssymb}
\usepackage{epsfig}
\usepackage{graphics}
\usepackage{subfigure}
\makeatletter
\def\eqalign#1{\null\,\vcenter{\openup\jot\m@th
  \ialign{\strut\hfil$\displaystyle{##}$&$\displaystyle{{}##}$\hfil
      \crcr#1\crcr}}\,}
\def\eqalignleft#1{\null\,\vcenter{\openup\jot\m@th
  \ialign{\strut$\displaystyle{##}$\hfil&$\displaystyle{{}##}$\hfil
      \crcr#1\crcr}}\,}
\makeatother 
\onecolumn
\begin{document}

\thesaurus{02 
            (02.01.2) 
           08 
            (08.02.1; 
             08.14.2)  
             13 
             (13.25.1) 
          }
\title{
New solution to viscous evolution of accretion disks in binary systems}

\author{ G.V.~Lipunova\inst{1} \and N.I.~Shakura\inst{1,2}
       }
\offprints{G.V.~Lipunova}
\mail{galja@sai.msu.ru}

\institute{Sternberg Astronomical Institute, Moscow State University,
        Universitetskii pr. 13,
        Moscow, 119899 Russia
        \and Max-Planck-Institut f\"ur Astrophysik, Karl-Schwarzschild-Str. 1,
        85740 Garching, Germany}

  \date{Received 31 May 1999 / Accepted 11 January 2000 }
 \maketitle

\begin{abstract}
Analytical investigation of time-dependent accretion in disks
is carried out. We consider a time-dependent disk in a binary system
at outburst which has a fixed tidally-truncated outer radius. The
standard Shakura-Sunyaev model of the disk is considered. The
vertical structure of the disk is accurately described in two regimes
of opacity: Thomson and free-free. Fully analytical solutions are
obtained, characterized by power-law variations of accretion rate with
time. The solutions supply asymptotic description of disk evolution
in flaring sources in the periods after outbursts while the disk is fully
ionized. The X-ray flux of multicolor (black-body) \hbox{$\alpha$-disk}
is obtained as varying quasi-exponentially. Application to X-ray
novae is briefly discussed concerning the observed faster-than-power
decays of X-ray light curves. The case of time-dependent advective disk
when the exponential variations of accretion rate can occur is discussed.
 \keywords{accretion, accretion disks --
 		binaries: close  --
    		novae, cataclysmic variables --
                X-rays: bursts
               }

   \end{abstract}

\newcommand{\marginlabel}[1]
             {\mbox{}\marginpar{\raggedleft\hspace{0pt}{\scriptsize\bf#1}}}
\newcommand{\de}{\partial}
\newcommand{\tc}{T_\mathrm{c}}
\newcommand{\omegak}{\omega_\mathsc{k}}
\newcommand{\nutc}{\nu_\mathrm{t}^\mathrm{c}}
\newcommand{\sigmasb}{\sigma_\mathrm{SB}}
\newcommand{\ls}{\left(}
\newcommand{\ps}{\right)}
\newcommand{\den}{\mathrm{d}}
\newcommand{\rout}{r_\mathrm{out}}
\newcommand{\kapt}{\varkappa_\mathsc{t}}
\newcommand{\taut}{\tau_\mathsc{t}}
\newcommand{\kapf}{\varkappa_\mathrm{ff}}
\newcommand{\tauf}{\tau_\mathrm{ff}}
\newcommand{\Bt}{B_\mathsc{t}}
\newcommand{\Bf}{B_\mathrm{ff}}
\newcommand{\mx}{m_\mathrm{x}}
\newcommand{\sigmo}{\Sigma_\mathrm{o}}
\newcommand{\tobs}{\Delta t_\mathrm{ob}}

\section{Introduction}

We investigate analytically the problem of time-dependent
accretion, closely related to the phenomena of flares widely
observed in binary systems.  In this paper we will concern ourselves with 
the emission of the flaring source which is 
generated by the accretion disk. We suppose the light curve to be
regulated by the accretion rate variations.  Such sources are typified by the
low-mass X-ray binaries and  cataclysmic variables.

 After Weizs\"acker~(\cite{Weiz48}) who considered
 the evolution of a  protoplanetary cloud, the analytical
 investigations of  non-stationary accretion
 were carried out by  L\"ust~(\cite{Lust52}), Lynden-Bell \&
 Pringle~(\cite{Lynd_prin74}), Lyubarskii \& Shakura~(\cite{Lyub_shak87},
 hereafter LS87) as applied to  accretion disks.
 A brief overview is presented in the book by Kato et
 al.~(\cite{Kato_etal98}).

 LS87 suggested three stages of evolution of a time-dependent accretion
 disk (see also \S~5.1). Initially a finite torus of the  increased
 density is formed around a gravitational centre. Viscosity causes the
 torus to spread and develop into the disk (1st stage). After the disk
 approaches the centre, the accretion rate reaches the maximum value
 (2nd stage) and begins to descend (3rd stage).
 During this stage the total angular momentum of the disk is conserved.
 
 Ogilvie~(\cite{Ogil99}) presented a time-dependent self-similar  analytical
 solution for a quasi-spherical advection-dominated flow with
 conserved total angular momentum.

 In a binary system, variations of accretion rate
 can be due to the non-stationary  exchange of mass between the
 components of the binary (mass-overflow instability model) or due
 to the disk instability processes
 (see Kato et al.~\cite{Kato_etal98} and references therein).
 At some time the accretion rate onto the centre  begins to augment.
We assume that the  maximum accretion rate through the inner boundary of
the disk corresponds to a peak of  outburst and the accretion rate
decreases afterwards.

In this study we particularly focus on the stage soon after the outburst.
In \S~2  we outline the general equations of time-dependent accretion.
The basic equation
relates surface density of the disk and viscous stresses in it. Thus the
specific structure of the disk influences greatly the run of the process.
\S~3 introduces the investigation of time-dependent Keplerian $\alpha$-disks.
The vertical structure of standard Keplerian disk is considered
in \S~4. In a binary system the  third  stage of LS87 cannot be realized
because the accretion disk around a primary would be confined by the
gravitational influence of a secondary. Such disks do not preserve their
angular momentum, transferring it to the orbital motion. We suggest the
particular conditions at the outer boundary of the disk which allow
the acquisition of new solutions characterized by faster decays than in LS87.
The procedure and the analytical solution are presented in \S~5.

We calculate the resulting bolometric light curve taking into
account the transition between the opacity regimes as accretion
rate decreases (\S~6). We note that observed light curves can have
different slopes due to unevenness of spectral distribution (\S~7).

In \S~8 we discuss the case of advection-dominated
accretion flow (ADAF) in which exponential 
variations with time of accretion rate possibly take place.

In \S~9 we discuss application of our model
to X-ray novae.

\section{Basic non-stationary accretion disk equation}

In the approximation of
Newtonian potential we assume that the velocity
of a free particle  orbiting at distance $r$
around a gravitating object is
\begin{equation}
\omegak r = (GM/r)^{1/2}~,
\label{kepler}
\end{equation}
where $\omegak$ is the Kepler angular velocity;
$M$ is the mass of the central gravitating  object, constant in time;
$G=6.67 \times 10^{-8}~$cm$^3\,$g$^{-1}\,$s$^{-1}\,$ is
the gravitational constant.
This is a good approximation to the law of motion for particles
in the standard sub-Eddington disk. In the advection-dominated
accretion flow (ADAF) the particles
are substantially subjected to the radial gradient of pressure
and thus  have the velocity different from that given by (\ref{kepler}).
Following the model by Narayan \& Yi~(\cite{Nara_yi94}),
one can assume that the
angular velocity in ADAF is $\omega = c_2 \,\omegak$~.

The height-integrated Euler equation on $\varphi$ and the
continuity equation along the height $Z$ are given by:
\begin{equation}
\sigmo\, v_r\,
\frac{\partial\, (\omega\,r^2)}{\partial r}
= - \,\frac{~1}{~r} ~\frac{\partial}{\partial r} \,(W_{r\varphi} r^2)~,
\label{1ur}
\end{equation}
\begin{equation}
\frac{\partial\sigmo}{\partial t} = -\,\frac{~1}{~r}\,
\frac{\partial}{\partial r}\, \sigmo v_r r~,
\label{2ur}
\end{equation}
where $\omega$ is the angular velocity in the disk;
\hbox{$\sigmo(r,t)=2\int\limits_0^{Z_\mathrm{o}}{\rho\, \mathrm{d}Z}$}
-- the surface density of the matter, and
$W_{r\varphi}(r,t)
= 2 \int\limits_0^{Z_\mathrm{o}}\, w_{r\varphi}\, \mathrm{d}Z$
is the height-integrated viscous shear stresses
between adjacent layers.  The time-independent angular
velocity is assumed although there can possibly be
certain variations of $\omega$
in the non-Keplerian advective disks when a time-dependent
pressure gradient is involved (see, e.g. Ogilvie~\cite{Ogil99}).

It is convenient to introduce the following variables:
$ F = W_{r \varphi}r^2$, henceforth $2\pi F$ means the total
moment of viscous forces acting between the adjacent  layers,
$h_\ast=\omega r^2$ -- the specific angular momentum of the matter
in the disk, and $h\equiv\omegak r^2$~.
From Eq.~(\ref{1ur}) in view of  (\ref{kepler}) it follows that
\begin{equation}
\sigmo\, v_r\, r~= \frac{\dot M (r,t)}{2\,\pi} =
-\,\left[ \frac{\partial h_\ast}{\partial h}\right]^{-1}\,
\frac{\partial F}{\partial h}~.
\label{Mdot}
\end{equation}
Substituting (\ref{Mdot}) in  (\ref{2ur}) and expressing
$r$ in terms of $h$, we obtain the basic
equation of time-dependent accretion:
\begin{equation}
\frac{\partial\sigmo}{\partial t} = \frac{1}{2}\, \frac{(GM)^2}{h^3}\,
\frac{\partial}{\partial h}
\left(\left[ \frac{\partial h_\ast}{\partial h}\right]^{-1}
      \frac{\partial F}{\partial h} \right)~.
\label{basic}
\end{equation}
In the case of the Keplerian disk $\partial h_\ast / \partial h=1$.
The advection-dominated solution by Narayan and Yi~(\cite{Nara_yi94})
yields $\partial h_\ast / \partial h=c_2$, where $c_2$ is a dimensionless
constant.

\section{Non-linear problem of evolution of the standard
Shakura-Sunyaev disk}

The special case when  the moment of viscous forces depends linearly
on the surface density and has a power law dependence on the radius
($F\propto \sigmo h^l$)
was thoroughly investigated by Lynden-Bell \& Pringle~(\cite{Lynd_prin74}).
In this particular case  Eq.~(\ref{basic}) is linear
and the solution can be presented as the superposition of
particular solutions (Green's functions) while the non-linear
equations do not allow such solutions.
In the paper by LS87 the necessary relation
between $\sigmo$ and $F$ for $\alpha$-disks (Shakura~\cite{Shak72};
Shakura \& Sunyaev~\cite{Shak_Suny73}) was derived from
the vertical structure equations.
Then Eq.~(\ref{basic})
acquires the following non-linear
form taking into account that $h\equiv h_\ast$:
\begin{equation}
\frac{\partial F}{\partial t}=
D\,\frac{F^m}{h^n}\,\frac{\partial^2F}{\partial h^2}~,
\label{nonlin}
\end{equation}
where  $D$ is the dimension constant;
$m=2/5$, $n=6/5$ when the Thomson scattering dominates the
opacity in the accretion disk, and
$m=3/10$, $n=4/5$ when the free-free and free-bound transitions do.
The ``diffusion constant'' $D$, defined  by the specific vertical structure,
relates $\sigmo$, $F$, and $h$:
\begin{equation}
\sigmo = \frac{(GM)^2\, F^{1-m}}{2\,(1-m)\,D\, h^{3-n} }
\label{SigDF}
\end{equation}
(see also Filipov~\cite{Fili84}).
$D$ is a function of $\alpha$, opacity coefficient,
and  the dimensionless values, which are the combinations of
the characteristic physical parameters of the disk. The value of $D$
is to be derived from the consideration of the disk vertical structure.
In the following section we obtain its value using the results
of the work by Ketsaris \& Shakura~(\cite{Kets_shak98}).

\section{Vertical structure of standard disk}

Hereafter, until specially mentioned, we assume that the matter in the
disk  moves with the Keplerian angular velocity $\omegak$,
and its state is governed by the ideal gas equation
\begin{equation}
P = \frac{\rho\,\Re\,  T}{\mu}~,
\label{ideal}
\end{equation}
where $\mu$ and $\Re=8.31\times 10^7$~erg~mol$^{-1}$K$^{-1}$ are
the molecular weight of the gas and the molar gas constant, respectively.
Along the $Z$ coordinate the hydrostatic equilibrium takes place:
\begin{equation}
\frac {1}{\rho} \,
\frac{\partial P}{\partial Z}
= - \omegak^2 Z~
\label{hydro}
\end{equation}
and the continuity equation is
\begin{equation}
\frac{\de \Sigma}{\partial Z} = \rho~.
\label{conti}
\end{equation}
We assume the radiation transfer equation in the diffusive
approximation:
\begin{equation}
\frac{c}{3\, \varkappa\, \rho}\, \frac{\partial (a T^4)}{\partial Z}
= -\, Q~,
\label{diffusive}
\end{equation}
where $c=2.99\times 10^{10}$~cm~s$^{-1}$
is the light velocity, $a=7.56 \times 10^{-15}$~erg~cm$^3$K$^4$.
The vertical gradient of the radiation flux $Q$
is proportional to the energy release per unit volume
in the disk; that is,
\begin{equation}
\frac{\partial Q}{\partial Z} = \varepsilon~~
[\mathrm{erg}\,\mathrm{cm}^{-3}\,\mathrm{s}^{-1}]~.
\label{flux}
\end{equation}
We take the opacities in the form of a power law
\hbox{$\varkappa =\varkappa_0 \rho^\zeta/T^\nu$} where
$\zeta=\nu=0$, $\varkappa_0=0.4$~cm$^2$g$^{-1}$
if \hbox{$\kapt$ $\gg$ $\kapf$} and
$\zeta=1$, $\nu=7/2$, $\varkappa_0=6.45\times 10^{22}$~cm$^5$K$^{7/2}$g$^{-2}$
if \hbox{$\kapf$ $\gg$ $\kapt$}.
Generally, in the optically thick disks  the energy release
can be described as a power law of temperature and density
(Tayler~\cite{Tayl80}). In a sense the calculation of the disk
structure resembles the calculation of stellar internal structure.
In the present study two cases are considered:  the energy release
$\varepsilon$ is proportional to (a) the pressure $\propto \rho\,T$,
		                 (b) the density $\rho$ alone. 
The thermal energy release is due to the
differential rotation of a viscous
disk:
\begin{equation}
\varepsilon = -\,r\, w_{r\varphi}\,\frac{\de\omega}{\de r}
= \frac 32 \,\omegak \,w_{r\varphi} ~.
\label{release}
\end{equation}
We follow Shakura~(\cite{Shak72}) and Shakura \& Sunyaev~(\cite{Shak_Suny73})
in suggesting that the turbulent viscous stress tensor is parameterized by
the pressure:
\begin{equation}
w_{r\varphi}=-\,\nu_\mathrm{t}\,\rho\,r\,\frac{\de\omega}{\de r}=
\frac 32\, \omegak\, \nu_\mathrm{t}\, \rho\, =
\alpha\, P ~,
\label{stress_mal}
\end{equation}
where $\nu_\mathrm{t}$ is the
kinematic coefficient of turbulent viscosity.
The height-integrated viscous stress tensor is given by
\begin{equation}
W_{r\varphi}(r,t) =2\, \int\limits_0^{Z_\mathrm{o}}w_{r\varphi}
\,  \mathrm{d}Z~
= 3\, \omegak\,\int\limits_0^{Z_\mathrm{o}}
\nu_\mathrm{t}\, \rho\, \mathrm{d}Z~.
\label{wrf}
\end{equation}
The energy emitted from the unit surface of one side of the
disk is obtained by integrating
(\ref{flux}) using (\ref{release}) and (\ref{wrf}):
\begin{equation}
Q_\mathrm{o} = \frac 12 \, W_{r\varphi}(r,t)\, r\,  \frac{\de\omega}{\de r}=
\frac 34\, \omegak \,    W_{r\varphi}(r,t)                  ~.
\label{qo}
\end{equation}

Above equations written for stationary accretion disks
hold in a non-stationary case
taking into account that the characteristic hydrostatic time of order
of $\alpha/ \omega$  is shorter than the time of radial movement
in the disk $r/v_r\sim (r/Z_\mathrm{o})^2/\alpha\omega $.

There are now various works investigating the
vertical structure of the disks. For example, Nakao
\& Kato~(\cite{Naka_kato95}) considered turbulent diffusion in the disk
providing the variations of viscous heating and $\alpha$-parameter
along the height $Z$.
The vertical structure of the disks including radiative and convective
energy transfer  was investigated by
Meyer \& Meyer-Hofmeister~(\cite{Meye_meye82}). They investigated two types
of viscosity, proportional to the gas pressure or to the total
pressure.  We adopt the result
of Ketsaris \& Shakura~(\cite{Kets_shak98}) who proposed a new method of calculating
the vertical structure of optically thick $\alpha$-disks assuming
power $\rho$-- and $T$-- dependences for the opacity and the energy release.

The dimensionless variable
\begin{displaymath}
\sigma = \frac{2\,\Sigma}{\sigmo}
\end{displaymath}
is introduced for convenient description of the problem, along
with the following variables: $p=P/P_\mathrm{c}\,$, $\theta=T/\tc\, $,
$z=Z/Z_\mathrm{o}\, $, $j=\rho/\rho_\mathrm{c}\, $,
and                    $q=Q/Q_\mathrm{o}\, $.
The method involves
the finding of the eigenvalues of the dimensionless parameters  in the
differential equations that describe vertical structure of the disk:
\footnote{Left bottom equation in formula (\ref{pppp}) is corrected 
in comparison with the journal variant}
\begin{equation}
\begin{array}{lll}
\displaystyle\frac{\mathstrut \mathrm{d}p}{\mathrm{d}\sigma} = &- \Pi_1 \,\Pi_2 \, z~;\qquad
&\Pi_1  = \displaystyle\frac{\mathstrut \omegak^2\, Z_\mathrm{o}^2\,\mu}{\Re\,T_\mathrm{c}}~;\\[5mm]
\displaystyle\frac{\mathstrut \mathrm{d}z}{\mathrm{d}\sigma} = &\Pi_2\, \displaystyle\frac{\theta}{p}~; \qquad
&\Pi_2  = \displaystyle\frac{\mathstrut \sigmo}{2\, Z_\mathrm{o}\, \rho_\mathrm{c}} ~; \\[5mm]
\displaystyle\frac{\mathstrut \mathrm{d}q}{\mathrm{d}\sigma} = &\Pi_3\, \theta ~; \qquad
&\Pi_3 = \displaystyle\frac{\mathstrut 3}{4}\,
\displaystyle\frac{\mathstrut \alpha\,\omegak\,\Re\,\tc\,\sigmo}{Q_\mathrm{o}\,\mu}~
\equiv \displaystyle\frac{\mathstrut \alpha\,\Re\,\tc\,\sigmo}{\mathstrut{W_{r\varphi}\,\mu}} ~; \\[5mm]
\displaystyle\frac{\mathstrut \mathrm{d}\theta}{\mathrm{d}\sigma} =
     &\Pi_4\, \displaystyle\frac{\mathstrut q\,j^{\zeta}}{\theta^{\nu+3}}   ~;  \qquad
&\Pi_4  = \displaystyle\frac{\mathstrut 3}{32}\,
           \left(\displaystyle\frac{\mathstrut T_\mathrm{ef}}{T_\mathrm{c}}\right)^4\,
          \displaystyle\frac{\mathstrut \sigmo\,\varkappa_0\,\rho_\mathrm{c}^{\zeta}}
{T_\mathrm{c}^{\nu}}~,
\label{pppp}
\end{array}
\end{equation}
using the definite boundary conditions in the disk. $T_\mathrm{c}$,
$\rho_\mathrm{c}$, $P_\mathrm{c}$ denote the values in the equatorial plane
of the disk and
$Q_\mathrm{o}= (ac/4)\,T_\mathrm{ef}^4$~.

After some algebraic manipulation of the right hand
equations in (\ref{pppp}) we obtain
$\sigmo$ written in terms of $W_{r\varphi}\,r^2$ and $\omega\,r^2$, which
in view of (\ref{SigDF}) yields:
\begin{equation}
D=\frac{1}{4(1-m)} \left\{
 \frac{2^{6+ \zeta+2\nu}\alpha^{8+\zeta+2\nu}}
 {\Pi_1^{\zeta}\,\Pi_2^{2\zeta}\,
 \Pi_3^{8+ \zeta+2\nu}\,\Pi_4^2 } \right.
 \left(\frac{\Re}{\mu}\right)^{8+2\nu}
 \left. \left(\frac{9  \varkappa_0 }
           {8 \, a\, c}\right)^2
(GM)^{12+8\zeta} \right\}^\frac{1}{10+3\zeta+2\nu}~,
\label{D1}
\end{equation}
where
\begin{displaymath}
m=\frac{4+2\zeta}{10+3\zeta+2\nu}\,,\qquad
n=\frac{12+11\zeta-2\nu}{10+3\zeta+2\nu}~.
\end{displaymath}
It is worth noting that $D$ depends on $\varkappa_0$ very weakly:
to a power of ${1/5}$ or ${1/10}$. This fact is believed to reduce
the effect of uncertainties in our knowledge of the real law of opacity.
The combination of $\Pi_{1,2,3,4}$ in (\ref{D1})  varies slightly
with the optical depth $\tau$, i.e. along the radius of the disk
(see Tables~\ref{ppppp_t},~\ref{ppppp_f}). Thus,
factor $D$  in the basic equation of time-dependent accretion
(\ref{nonlin}) is considered to be constant.

Specific energy dissipation $\varepsilon/\rho\,=\,\de Q/\de \Sigma$
is defined by the temperature variations over $Z$. In principle, the
intensive stirring in the disk can account for
the situation when the  energy release per unit mass does not depend
on the height $Z$. This refers to the case (b) mentioned above where
$\varepsilon$ is the function of density.
In this situation the temperature dependence disappears in the
energy production equation (third line of (\ref{pppp})) and $\Pi_3=1$.

Ketsaris \& Shakura~(\cite{Kets_shak98}) calculated the values of $\Pi_1, \Pi_2, \Pi_3$,
and $\Pi_4$. Selected values  of $\Pi_{1,2,3,4}$  and corresponding
values of $\delta$, in the Thomson opacity regime,
and some effective optical thickness of the disk
$\tau_0 = \sigmo\, \varkappa_0\, \rho_\mathrm{c}/(2\,\tc^{7/2})$,
in the free-free regime,  are shown in Tables~\ref{ppppp_t},~\ref{ppppp_f}.
For the full version of $\Pi_{1,2,3,4}$ list and discussion
the reader is referred to the original paper by  Ketsaris \&
Shakura~(\cite{Kets_shak98}).
The parameter $\delta$ was introduced by them
for the sake of convenience and denotes the ratio
of total scattering optical thickness $\kapt\, \sigmo $ to that at the
thermalization depth\footnote{Formula (\ref{delta})
is corrected in comparison with the journal variant}:
\begin{equation}
\delta = \frac{\kapt\, \sigmo/2}{\taut(\tau^* =1)}~,\qquad
\tau^* = \int\limits_{Z^*}^{Z_\mathrm{o}} (\kapf\,\kapt)^{1/2}\,\rho\,
\mathrm{d} Z~,
\label{delta}
\end{equation}
where $ \tau^*$ is the effective optical depth (Zeldovich \&
Shakura~\cite{Zeld_shak69}; Mihalas~\cite{Miha78}).

\renewcommand{\arraystretch}{0.9}
\begin{table}
\caption[]{Vertical structure parameters in the Thomson opacity regime}
\label{ppppp_t}
\begin{tabular}{lllll|lll}
\hline
&&&&&\multicolumn{3}{c}{$\Pi_3 =1$}\\
\cline{6-8}
$\log{\delta}$&$\Pi_1$&$\Pi_2$&$\Pi_3$&$\Pi_4$& $\Pi_1$&$\Pi_2$&$\Pi_4$\\
\hline
4.00& 6.37&0.516&1.150 &0.460&6.46&0.511&0.500\\
3.00& 5.67&0.546&1.149 &0.459&5.74&0.542&0.499\\
2.00& 4.47&0.610&1.142 &0.454&4.52&0.605&0.490 \\
1.00& 2.61&0.740&1.105 &0.398&2.63&0.737&0.417\\
\hline
\end{tabular}
\end{table}

\renewcommand{\arraystretch}{0.9}
\begin{table}
\caption[]{Vertical structure parameters in the free-free opacity regime} 
\label{ppppp_f}
\begin{tabular}{lllll|lll}
\hline
&&&&&\multicolumn{3}{c}{$\Pi_3 =1$}\\
\cline{6-8}
$\log{\tau_0}$&$\Pi_1$&$\Pi_2$&$\Pi_3$&$\Pi_4$& $\Pi_1$&$\Pi_2$&$\Pi_4$\\
\hline
4.00& 7.07&0.487&1.131 &0.399&7.12&0.485&0.437\\
3.00& 6.31&0.515&1.131 &0.398&6.34&0.514&0.436\\
2.00& 4.98&0.576&1.126 &0.395&4.98&0.576&0.431 \\
1.00& 2.83&0.716&1.095 &0.354&2.81&0.716&0.373\\
\hline
\end{tabular}
\end{table}

\section{Time-dependent accretion in Keplerian disk}

\subsection{Solutions to non-stationary
Keplerian disk equation}

The self-similar solutions of Eq.~(\ref{nonlin}) were found by
LS87.
In these solutions
any physical characteristic of the disk, for instance, the surface density
$\sigmo(r,t)$, can be presented in the form:
$\sigmo(r,t)=S(t)\, s(r/R(t))$, where the scales $S(t)$ and $R(t)$
depend on $t$ in a particular way, and $s(r/R(t))$ is
a universal function of
one self-similar variable $\hat\xi=r/R(t)$  (Zeldovich \&
Raizer~\cite{Zeld_raiz67}).
The solutions represent three stages of the non-stationary accretion on
an object. The first stage is the formation
of the accretion disk from some finite torus around an object. The second
stage is the de\-ve\-lo\-ping of the quasi-stationary regime of accretion,
and the third --
the decay of accretion when the external boundary of the
disk is spreading away to infinity. LS87
obtained the self-similar solutions of type II for the first two stages and
the self-similar solution of type I (Zeldovich \& Raizer~\cite{Zeld_raiz67})
for the final stage (when there is conservation of the total angular
momentum of the disk).

In a binary system the accretion picture
has particular features. The main feature is
the limitation of the outer radius.
Thus, one cannot apply the LS87 solution at the third stage,
that is, during the decay of accretion after the peak of the outburst.
The spreading of the
disk is to be confined by the tidal interactions. The tidal
torque produced by a secondary star has strong
radial dependence (Papaloizou \& Pringle~\cite{Papa_prin77}). As Ichikawa \&
Osaki~(\cite{Ichi_osak94}) showed,
the tidal effects are generally small in the
accretion disk, except near to the tidal  truncation radius, which
is given by the last non-intersecting periodic particle orbit in the disk
(Paczy\'nski~\cite{Pacz77}). They concluded 
that once the disk expands to the tidal truncation
radius, the tidal torques prevent the disk from expanding beyond the
tidal radius.

A class of solutions of Eq.~(\ref{nonlin}), which this paper
focuses on, can be found  on separating
the variables $h$ and $t$.
We seek the solution  in the form $F(h,t)=F(t)f(\xi)$, where
$\xi=h/h_\mathrm{o}$, $h_\mathrm{o} = (GMr_\mathrm{out})^{1/2}$~.
From Eq.~(\ref{Mdot}), substituting $h\equiv h_\ast $, it
follows that
\begin{equation}
\dot M(h,t) = -2\pi f'(h/h_\mathrm{o}) F(t)/h_\mathrm{o}~.
\label{Mdotsol}
\end{equation}
Substitution of the product of two functions in
Eq.~(\ref{nonlin}) gives the time-dependent part of the solution:
\begin{equation}
F(t) = \left( \frac{h_\mathrm{o}^{n+2}}
           {-\lambda\, m\, D\, (t+t_0)} \right) ^ {1/m}~.
\label{F(t)SS}
\end{equation}
$D$ is the constant defined by the vertical structure of the disk
(\S~4, Eq. (\ref{D1}));
$\lambda$ is a negative separation constant which can be found from
boundary conditions on $f(\xi)$;
$t_0$ is an integration constant. From here on we set $t_0=0$ for the
Thomson opacity regime. We calculate a value of $t_0$ for the free-free
opacity regime in \S~6.2. Expression (\ref{F(t)SS}) represents asymptotic
law for after-peak evolution of a real source.

The equation for $f(\xi)$ is a non-linear differential equation of
second order which is a particular case of the general
Emden-Fowler equation (Zaycev \& Polyanin~\cite{Zayc_poly96}):
\begin{equation}
\frac{d^2 f}{d \xi^2} = \lambda \xi^n f^{1-m}~,
\label{f(h)SS}
\end{equation}
the solution of which we seek as a polynomial
\begin{equation}
f(\xi) = a_0\xi + a_1 \xi^k +a_2 \xi^l+ \dots~.
\label{f(xi)}
\end{equation}
Substituting $f(\xi)$ into Eq.~(\ref{f(h)SS}) we obtain for
the second and the third term:
\begin{equation}
\eqalignleft {
&k=3+n-m, \quad a_1=\frac{\lambda a_0^{1-m}}{k(k-1)}~, \cr
&l=2k-1,  \quad a_2=\frac{\lambda a_0^{-m}a_1}{l(l-1)} (1-m)~,
}
\label{SolSS}
\end{equation}
and $a_0$, $\lambda$ are to be defined from the boundary
conditions on $f(\xi)$.

\begin{figure}
\begin{center}
\rotatebox{-90}{{\resizebox{4.5cm}{!}{\includegraphics{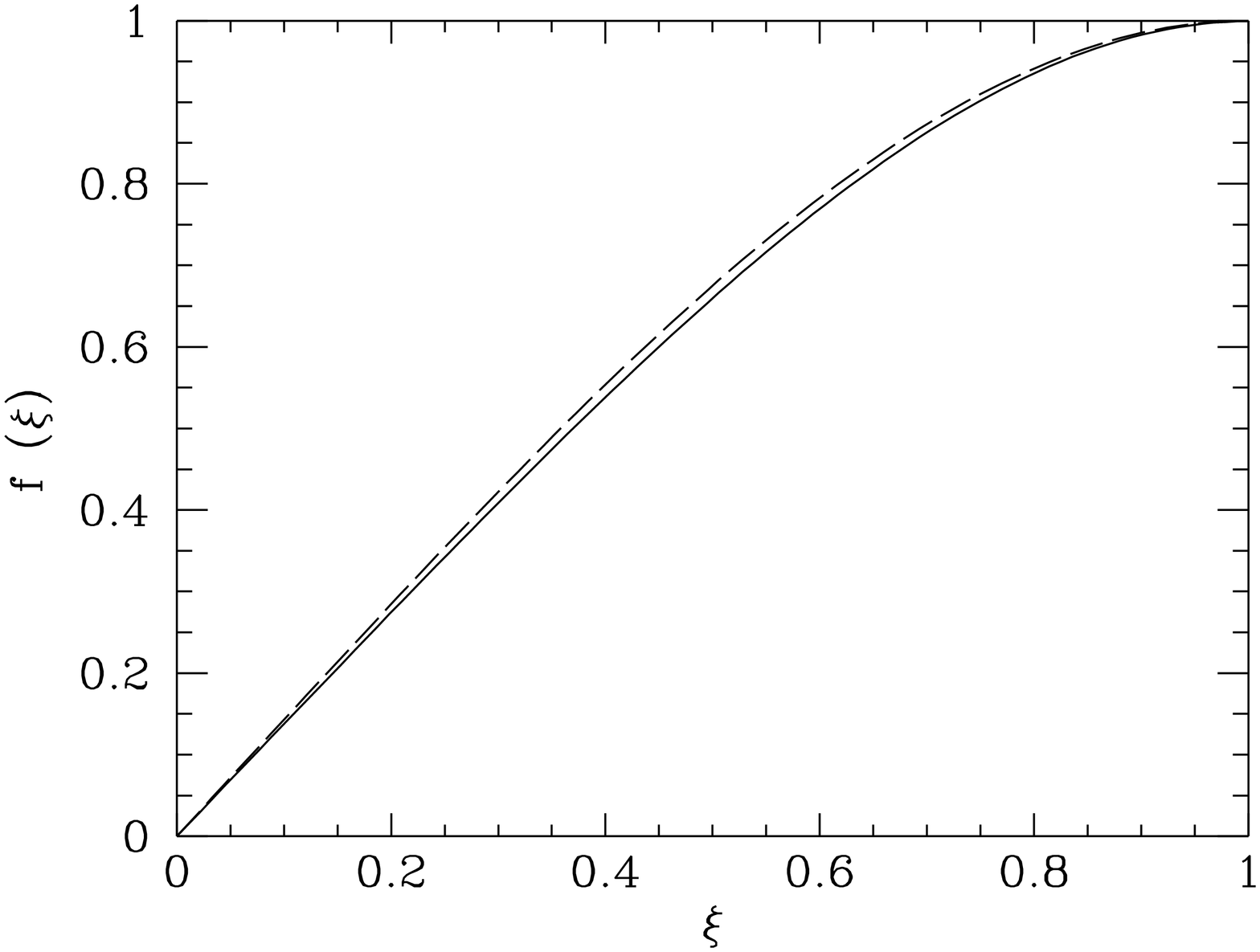}}}}
\end{center}
\vskip -0.5 cm
 \caption{
 The solution $f(\xi)$ in two cases: when $\kapt$ $\gg$ $\kapf$ (solid line)
  and  \hbox{$\kapf$ $\gg$ $\kapt$} (dashed line)
 }
 \label{fshtrih}
\end{figure}

We consider the size of the disk to be maximum and invariant
over the period of outburst.
As the drain of angular momentum occurs in a narrow region near this
truncation radius (Ichikawa \& Osaki~\cite{Ichi_osak94}),  we treat the region
near this radius
as the $\delta$-type channel, not considering the details of the process.
In other words, the smooth behaviour of spatial factor $f$ in the moment
of viscous forces $F$ (which increases
as $\propto r^{1/2}$ in the inner parts of the disk, then flattens, reaches
the maximum and drops down near $\rout$ due to tidal torque)
is  analytically treated as
increasing, flattening, and reaching maximum at $\rout$, which is the
end of the disk (this profile is shown in Fig.~\ref{fshtrih}).
Thus we propose the boundary conditions as follows:
\begin{equation}
f(1)=1, \quad  f'(1)=0~.
\label{bound}
\end{equation}
Corresponding $a_0$ and  $\lambda$ are displayed in Table \ref{lam_a0}.

Naturally, real accretion disks have finite value of $r_\mathrm{in}\neq 0$,
but still, in most cases,
$r_\mathrm{in}/r_\mathrm{out} \ll 1$, that is equivalent to
$r_\mathrm{in}/r_\mathrm{out} = 0$ in our problem from the mathematical standpoint.

\renewcommand{\arraystretch}{0.9}
\begin{table}
\caption{Summary of parameters in solutions
for two opacity regimes for the Keplerian disk}
\label{lam_a0}
\begin{tabular}{lllllllll}
\hline
                      &m      &n      &$\lambda$& $a_0$   &$a_1$ &$a_2$&$k$&$l$  \\
\hline
$\kapt$ $\gg$ $\kapf $& $2/5$ & $6/5$ &$-3.482$ & $1.376$ &$-0.39$ &0.02 &3.8&6.6  \\
$\kapf$ $\gg$ $\kapt$ & $3/10$&$4/5$  &$-3.137$ & $1.430$ &$-0.46$ &0.03 &3.5&6.0  \\
\hline
\end{tabular}
\end{table}
Note that (\ref{F(t)SS}) implies a considerably
{\em steeper} time dependence 
than the solution by LS87 does. The latter yields the accretion rate as a
function of $t^{-19/16}$ if $\kapt$ $\gg$ $\kapf $, and
$ t^{-5/4}$ if $\kapf$ $\gg$ $\kapt$.  In our case this dependence
is  $ t^{-5/2}$   if $\kapt$ $\gg$ $\kapf $,  and $ t^{-10/3}$
if $\kapf$ $\gg$ $\kapt$.
This difference is due to the non-conservation
of angular momentum in the disk in our case.

The following subsections contain the explicit
expressions for the  physical characteristics
of the disk. They are deduced from
(\ref{SigDF}), (\ref{pppp}), (\ref{D1}), and (\ref{F(t)SS}).
We introduce for the mass of the central object the
quantity $\mx = M/M_{\sun}$.

\subsection{Thomson opacity regime ($\kapt$ $\gg$ $\kapf$)}
Here the function $f=f(\xi)=f((r/\rout)^{1/2})$ and the
values $\Pi_{1..4}$ should be taken for the Thomson
opacity regime.
Then we have\footnote{Formulae (\ref{d_tomson}) and (\ref{z1})
are corrected in comparison with the journal variant}:
\begin{equation}
D~[\mathrm{g}^{-2/5}\, \mathrm{cm}^{28/5}\,\mathrm{s}^{-17/5}]~=
2.42\times 10^{38}\,~
 \alpha^{4/5}\, \mx^{6/5}\, \ls\frac{\mu}{0.5}\ps^{-4/5}\,
\ls{\Pi_3^4\,\Pi_4}\ps^{-1/5} ~,
\label{d_tomson}
\end{equation}
\begin{equation}
\sigmo~[\mathrm{g}\, \mathrm{cm}^{-2}]=
3.4\times 10^{2}\,~
\alpha^{-2}\, \mx^{1/2}\,
\ls\frac{\mu}{0.5}\ps^{2}\,
\ls\frac{r}{\rout}\ps^{-9/10}\, f^{3/5}\,
\ls\frac{\rout}{R_{\sun}}\ps^{3/2}\,
\ls\frac{t}{10^\mathrm{d}}\ps^{-3/2}
\ls{\Pi_3^4\,\Pi_4}\ps^{1/2}~,
\end{equation}
\begin{equation}
\tc~[\mathrm{K}]=1.8\times 10^{4}\,
\alpha^{-1}\, \mx^{1/2}\,
\ls\frac{\mu}{0.5}\ps \,
\ls\frac{r}{\rout}\ps^{-11/10}\, f^{2/5}\,
\ls\frac{\rout}{R_{\sun}}\ps^{1/2}\,
\ls\frac{t}{10^\mathrm{d}}\ps^{-1}\,
\Pi_3 \,,
\label{Tomsony}
\end{equation}
\begin{equation}
\frac{z_\mathrm{o}}{r}
= 0.04~
\alpha^{-1/2}\,\mx^{-1/4}\,
\ls\frac{r}{\rout}\ps^{-1/20}\, f^{1/5}\,
\ls\frac{\rout}{R_{\sun}}\ps^{3/4}\,
\ls\frac{t}{10^\mathrm{d}}\ps^{-1/2}\,
\ls {\Pi_1\,\Pi_3}\ps^{1/2}~,
\label{z1}
\end{equation}
\begin{equation}
\tau^*=4.8\times 10^{2}~ \alpha^{-1}\,
\ls\frac{\mu}{0.5}\ps^{5/4} \,
\ls\frac{r}{\rout}\ps^{1/10}\, f^{1/10}\,
\ls\frac{\rout}{R_{\sun}}\ps^{1/2}\,
\ls\frac{t}{10^\mathrm{d}}\ps^{-1/4}\,
\ls\frac{\Pi_3^4\,\Pi_4^3}{\Pi_1\,\Pi_2^2}\ps^{1/4}~.
\end{equation}
$ \tau^*$ is the effective optical thickness of the disk defined
by the combined processes of scattering and absorption. We
take approximately (c.f. (\ref{delta})):
$$
\tau^* =   \left(\frac{0.4\times 6.45\times 10^{22}~
\rho_\mathrm{c}}{\tc^{7/2}} \right)^{1/2}~\sigmo~.
$$

\subsection{Free-free opacity regime ($\kapf$ $\gg$ $\kapt$)}
Here the function $f$
and the
values $\Pi_{1..4}$ should be taken for the free-free opacity regime.
The following formulae contain the constant $t_0$ appeared in 
expression (\ref{F(t)SS}). It was  neglected in the previous subsection;
here $t_0$
accounts for the possibility of time shifts between the solutions
in the two opacity regimes.
We have\footnote{Formulae (\ref{d_freefree}) and (\ref{z2}) are corrected
 in comparison with the journal variant}:
\begin{equation}
D~[\mathrm{g}^{-3/10}\, \mathrm{cm}^{5}\,\mathrm{s}^{-16/5}]
=5.04\times 10^{34}\,
\alpha^{4/5}\, \mx\, \ls\frac{\mu}{0.5}\ps^{-3/4}\,
\ls{\Pi_1^{1/2}\, \Pi_2\,\Pi_3^8\,\Pi_4}\ps^{-1/10}\,,
\label{d_freefree}
\end{equation}
\begin{equation}
\sigmo~[\mathrm{g}\, \mathrm{cm}^{-2}]
=5.3\times 10^{2}\,
\alpha^{-8/3}\, \mx^{5/6}\,
\ls\frac{\mu}{0.5}\ps^{5/2}\,
\ls\frac{r}{\rout}\ps^{-11/10}\, f^{7/10}\,
\ls\frac{\rout}{R_{\sun}}\ps^{13/6}\,
\ls\frac{t+t_0}{10^\mathrm{d}}\ps^{-10/3}
\ls{\Pi_1^{1/2}\, \Pi_2\,\Pi_3^8\,\Pi_4}\ps^{1/3}\!\!\!\!,
\end{equation}
\begin{equation}
\tc~[\mathrm{K}]=3.1\times 10^{4}\,
\alpha^{-1}\, \mx^{1/2}\,
\ls\frac{\mu}{0.5}\ps \,
\ls\frac{r}{\rout}\ps^{-9/10}\, f^{3/10}\,
\ls\frac{\rout}{R_{\sun}}\ps^{1/2}\,
\ls\frac{t+t_0}{10^\mathrm{d}}\ps^{-1}\,
{\Pi_3}~,
\label{frees}
\end{equation}
\begin{equation}
\frac{z_\mathrm{o}}{r}
= 0.05 ~
\alpha^{-1/2}\,\mx^{-1/4}\,
\ls\frac{r}{\rout}\ps^{1/20}\, f^{3/20}\,
\ls\frac{\rout}{R_{\sun}}\ps^{3/4}\,
\ls\frac{t+t_0}{10^\mathrm{d}}\ps^{-1/2}\,
\ls{\Pi_1\,\Pi_3}\ps^{1/2}~,
\label{z2}
\end{equation}
\begin{equation}
\tau=4.5\times 10^{2}~  \alpha^{-4/3}\,\mx^{1/6}\,
\ls\frac{\mu}{0.5}\ps^{3/2} \,
\ls\frac{r}{\rout}\ps^{-1/10}\, f^{1/5}\,
\ls\frac{\rout}{R_{\sun}}\ps^{5/6}\,
\ls\frac{t+t_0}{10^\mathrm{d}}\ps^{-2/3}\,
\ls\frac{\Pi_3^4\,\Pi_4^2}{\Pi_1^{1/2}\,\Pi_2}\ps^{1/3}~.
\label{tau_free}
\end{equation}
This regime is characterized by lower temperature and density, and
the optical thickness of the disk is defined by the
processes of free-free absorption:
$\tau = 2\,\tau_0 =  6.45\times 10^{22}\,
\rho_\mathrm{c}\,\tc^{-7/2}\,\sigmo\,$.

\section{Bolometric light curves of time-dependent standard accretion disk:
power law}

In order to calculate the luminosity of the disk, we assume
the quasi-stationary accretion rate as it is at $r\ll\rout$.
For these most luminous parts of the disk we take
$\dot M(t)=\dot M(0,t)$ given by (\ref{Mdotsol}). 
The
overall emission of the disk is  defined by the gravitational energy release
$L = \eta\, \dot M(t)\, c^2$~, where $\eta$ is the efficiency of the process.

At early $t$, when the Thomson scattering is
dominant, we derive that the bolometric luminosity of the disk
varies as follows:
\begin{equation}
L_\mathrm{T}(t)~[\mathrm{erg~s}^{-1}]
= 1.3\times 10^{39}~
\alpha^{-2}\,  m_\mathrm{x}^{1/2}\,
\ls\frac{\eta}{0.1} \ps \ls\frac{\mu}{0.5}\ps^2\,
\ls\frac{r_\mathrm{out}} {R_{\sun}}\ps^{7/2}\,
\ls\frac{t}{10^\mathrm{d}}\ps^{-5/2}\,
\ls{\Pi_3^4\, \Pi_4}\ps^{1/2}~.
\label{Lbol_t}
\end{equation}
As the temperature decreases, the law of decline switches to:
\begin{equation}
L_\mathrm{ff}(t)~[\mathrm{erg~s}^{-1}] =  3.6\times 10^{39}\,
\alpha^{-8/3}\,  m_\mathrm{x}^{5/6}\,
\ls \frac{\eta}{0.1} \ps \ls\frac{\mu}{0.5}\ps^{5/2}\,
\ls\frac{r_\mathrm{out}} {R_{\sun}}\ps^{25/6}\,
\ls\frac{t+t_0}{10^\mathrm{d}}\ps^{-10/3}\,
\ls{\Pi_1^{1/2}\, \Pi_2\,\Pi_3^8\,\Pi_4}\ps^{1/3}        ~.
\label{Lbol_f}
\end{equation}
The mass of the disk can be derived by integrating $\sigmo$ over its
surface:
\begin{equation}
M_\mathrm{disk,T}~[M_{\sun}]=4 \times 10^{-9}   \,     \alpha^{-2} \,
{m_\mathrm{x}^{1/2} \, \ls \frac{\mu}{0.5}\ps^2  \,
\ls\frac{\rout}{R_{\sun}}\ps^{12/5} }\,
{\ls \frac{t}{10^\mathrm{d}}\ps^{-3/2}\,\ls{\Pi_3^4\,\Pi_4}\ps^{1/2},        }
\label{Mtot_t}
\end{equation}
\begin{equation}
M_\mathrm{disk,ff}~[M_{\sun}]= 2\times 10^{-8}\,  \alpha^{-8/3} \,
{m_\mathrm{x}^{5/6} \, \ls \frac{\mu}{0.5}\ps^{5/2}  \,
\ls\frac{\rout}{R_{\sun}}\ps^{49/15} }\,
{\ls \frac{t+t_0}{10^\mathrm{d}}\ps^{-7/3}\,\ls{\Pi_1^{1/2}\, \Pi_2\,\Pi_3^8\,\Pi_4}\ps^{1/3}  }~.
\label{Mtot_f}
\end{equation}
The constant $t_0$ is the same as in the previous section.
These solutions give an asymptotic law for the disk
{\em  bolometric} luminosity variations.
The value of $t_0$ will be obtained in \S~6.2
when we shall discuss the transition between the regimes
of opacity.

We remark that the observed X-ray
light curves can have {\em different} (most probably, steeper)
law of decay. Indeed, the
energy band of an X-ray detector usually covers the region harder
$1$~keV where the multi-color photon spectrum of the
disk (having appropriate temperature) can have turnover
from  $-2/3$ power law into exponential fall. This turnover
is expected to change its position due to variations in temperature
of the disk after the burst. The narrower the
observed band, the more different the
observed curve could look like in comparison with the  expected bolometric
flux light curve. In \S~7 we discuss this subject in more detail.

\subsection{Luminosity -- accretion disk parameters dependence}

It is essential to point out that in formulae (\ref{Lbol_t}), (\ref{Lbol_f})
the parameters $\alpha, \mu, m, r_\mathrm{out} , \Pi_{1,2,3,4}$
cannot be changed to describe how luminosity depends on them.
Indeed, these expressions were found as a result of solution of 
differential Eq.~(\ref{nonlin}) with the constant coefficient $D$
(which depends on parameters of the disk, except $\eta$). Imagine a 
situation when
one of these parameters, say $\alpha$, quickly
increases. This will not result in the decrease of the luminosity
as it might seem from (\ref{Lbol_t}) or (\ref{Lbol_f}).
What will happen really is that the accretion will change to another
solution (during the same regime of opacity),
according to the new $D^*$.
Supposing that the mass of the disk remains constant during this transition
and taking into account that the profile of $\sigmo(r)$ does not change,
it can be seen from  (\ref{SigDF}) that  $F$ changes discontinuously
and  the luminosity $L\propto\dot M\propto F(t)$
jumps   as $(\alpha^*/\alpha)^{4/(5(1-m))}$,
$\alpha^* > \alpha$.
The relation between the new and the old $F$ and $D$, obtained
from (\ref{SigDF}), gives the new term $(t+t_0^*)$ in  (\ref{F(t)SS}):
\begin{equation}
\frac{t+t_0^*}{t+t_0} = \ls\frac{D}{D^*}\ps^{1/(1-m)}~.
\label{t*t}
\end{equation}
Thus the increase in $\alpha$ gives the increase in $D$ and, consequently,
$(t+t_0^*) < (t+t_0)$ which implies a steeper light curve after
the transition than before. The increase of $\alpha$ can be possibly
provided by the enhanced role of convection
in the accretion disk and will result in the brightening of the disk.
This situation is displayed in the inset in Fig.~\ref{lumin}. We note
that the descending portion of the curve after the increase is uncertain
if convection is involved since the disk structure modifies from that
presented in \S~4.

\subsection{Thomson opacity -- free-free opacity transition}
The temperature of the disk decreases with time, and eventually
the free-free and free-bound opacity supersedes the Thomson one.
It is possible to connect two regimes at the point
\hbox{($\xi$, $t_\mathrm{tr}$)},
where $F_1(\xi,t_\mathrm{tr})=F_2(\xi,t_\mathrm{tr}+t_0)$
and ${\sigmo}_{,1} = {\sigmo}_{,2}$
(indexes 1, 2 denote different opacity regimes) -- two conditions allowing us
naturally to define both $t_\mathrm{tr}$ and $t_0$:
\begin{equation}
\begin{array}{l}
t_\mathrm{tr} = 3.7^\mathrm{d} \,
 m_\mathrm{x}^{2/5}\, \alpha^{-4/5}\,
\ls\displaystyle\frac{\mathstrut \mu}{0.5}\ps^{3/5}\,
\ls\displaystyle\frac{\mathstrut r}{\rout}\ps^{-4/5}\,
\ls\displaystyle\frac{\mathstrut \rout}{R_{\sun}}\ps^{4/5}\,
\displaystyle\frac{\mathstrut ^\mathrm{f}f^{12/5}}{^\mathsc{t}f^{2}}\,
\ls\displaystyle\frac{\mathstrut ^\mathrm{f}\Pi_1\,^\mathrm{f}\Pi_2^2\,
^\mathrm{f}\Pi_3^{16}\, ^\mathrm{f}\Pi_4^2}
{^\mathsc{t}\Pi_3^{12}\,^\mathsc{t}\Pi_4^4}\ps^{1/5}~,  \\[5mm]
t_\mathrm{tr}+t_0 = 6.4^\mathrm{d} \,
 m_\mathrm{x}^{2/5}\, \alpha^{-4/5}\,
\ls\displaystyle\frac{\mathstrut \mu}{0.5}\ps^{3/5}\,
\ls\displaystyle\frac{\mathstrut r}{\rout}\ps^{-3/5}\,
\ls\displaystyle\frac{\mathstrut \rout}{R_{\sun}}\ps^{4/5}\,
\displaystyle\frac{\mathstrut ^\mathrm{f}f^{21/10}}{^\mathsc{t}f^{9/5}}\,
\ls\displaystyle\frac{\mathstrut ^\mathrm{f}\Pi_1\,^\mathrm{f}\Pi_2^2\,
^\mathrm{f}\Pi_3^{16}\, ^\mathrm{f}\Pi_4^2}
{^\mathsc{t}\Pi_3^{12}\,^\mathsc{t}\Pi_4^4}\ps^{1/5}.
\end{array}
\end{equation}
The right top indexes of $f= f(\xi)$
and $\Pi_{1,2,3,4}$ indicate the opacity regimes.
 As the profiles of $f(\xi)= f((r/\rout)^{1/2}) $
are very close in these two regimes (see Fig.~\ref{fshtrih}),
and parameters $\Pi_{1,2,3,4}$
vary slightly with radius (being roughly constant in the
region
where the substantial mass of the disk is enclosed),
the physical parameters of the disk ($\sigmo(r)$, $\tc(r)$, etc.) calculated
in the two solutions are sufficiently accurately equal.

At the time $t_\mathrm{tr}$  the free-free absorption coefficient
$\kapf=\varkappa_0 \rho_\mathrm{c}^\zeta/T_\mathrm{c}^\nu$
calculated in the Thomson opacity regime and in the free-free opacity regime
takes the form:
\begin{equation}
\begin{array}{l}
^\mathrm{f}\kapf~[\mathrm{cm}^2 \mathrm{g}^{-1}] =
0.399 \,  \ls\displaystyle\frac{\mathstrut ^\mathrm{f}f}{ ^\mathsc{t}f}\ps^6\,
\displaystyle\frac{\mathstrut ^\mathrm{f}\Pi_1^{1/2}\,^\mathrm{f}\Pi_2\, ^\mathrm{f}\Pi_3^{8}\,
^\mathrm{f}\Pi_4}
{^\mathsc{t}\Pi_1^{1/2}\,^\mathsc{t}\Pi_2\, ^\mathsc{t}\Pi_3^{8}\,
^\mathsc{t}\Pi_4}~, \\[5mm]
^\mathsc{t}\kapf~[\mathrm{cm}^2 \mathrm{g}^{-1}] =
0.399 \,  \ls\displaystyle\frac{\mathstrut ^\mathrm{f}f}{ ^\mathsc{t}f}\ps^3\,
\displaystyle\frac{\mathstrut ^\mathrm{f}\Pi_3^{4}\,^\mathrm{f}\Pi_4}
     {^\mathsc{t}\Pi_3^{4}\,^\mathsc{t}\Pi_4}~.
\end{array}
\end{equation}
The closeness of $\kapf$  to  $\kapt=0.4$~cm$^2$g$^{-1}$
confirms the reliability of our calculations
and yields the smoothness of the transition.

\begin{figure}
\begin{center}
\vbox to -2.7cm{
\vbox to 3mm{\vss
\hbox to\textwidth{
\hspace{8.8cm}
\resizebox{3.5cm}{!}{\rotatebox{-90}{\includegraphics{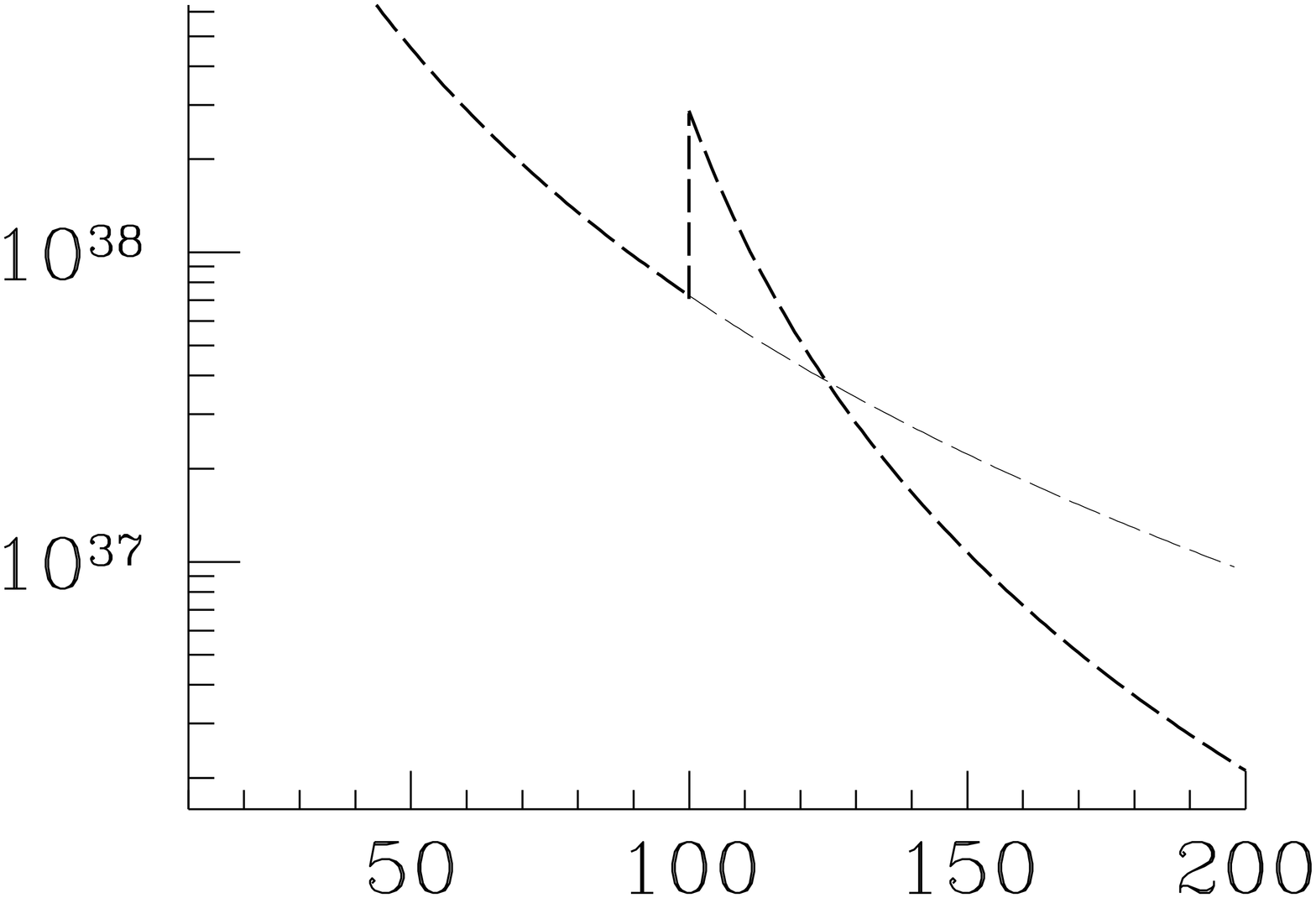}}}\hss}}}
\centerline{\resizebox{9cm}{!}{\rotatebox{-90}{\includegraphics{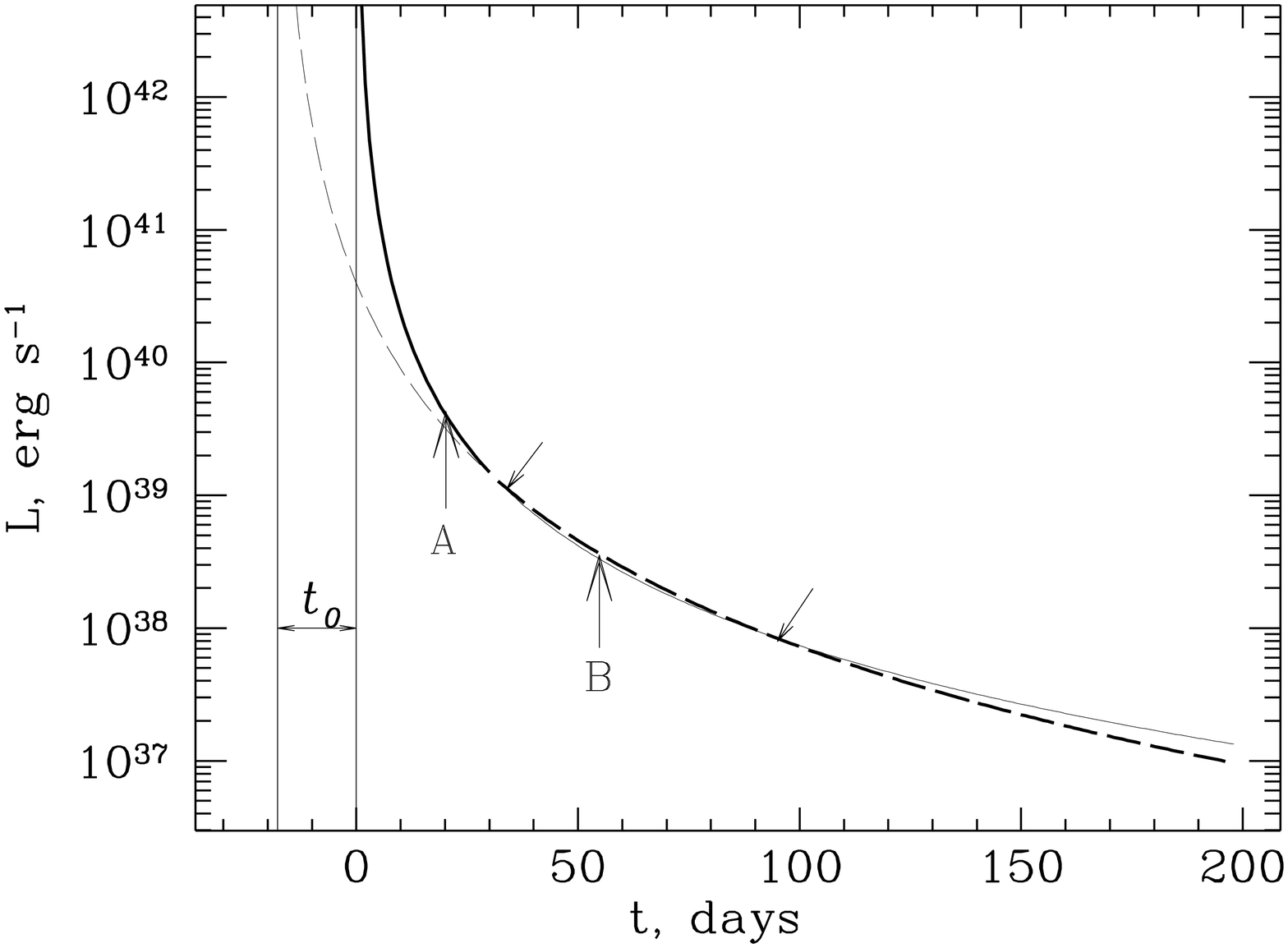}}}}
\end{center}
\vskip -0.5 cm
\caption{
Bolometric luminosity $ L_\mathrm{T}$ and $L_\mathrm{ff}$
calculated for  parameters:
 $\mx=3$, $\alpha=0.3$, $\mu=0.5$, $\rout=R_{\sun}$.
 Shown are the solution in Thomson opacity regime (solid line)
 and in the free-free opacity regime (dashed line).
 Their bold parts represent the resulting light curve
 of the disk.
 Small arrows mark two intersections when  $F_1(\xi,t)=F_2(\xi,t+t_0)$.
 The inset illustrates the case of increase of $\alpha$ from
 $0.3$ to $1$
  }
 \label{lumin}
\end{figure}

Fig. \ref{lumin} represents the bolometric light curve of the disk for
$\alpha=0.3$, $m_\mathrm{x}=3$. Hereafter we substitute $\Pi_{1,2,3,4}$
with their typical values in a self-consistent way.
The transfer between Thomson and free-free regimes
begins at the moment $r/\rout=\xi^2=1$,
$t\approx 8 ^\den (m_\mathrm{x}/3)^{2/5} \,\alpha^{-4/5}\,(\mu/0.5)^{3/5}\,
 (\rout/R_{\sun})^{4/5}$ -- arrow A at $21^\den$ in Fig. \ref{lumin}.
We intersect the curves  at
\begin{displaymath}
r/\rout=0.5, \quad
t=t_\mathrm{tr}\approx 13 ^\den (m_\mathrm{x}/3)^{2/5} \,
\alpha^{-4/5}\,(\mu/0.5)^{3/5}\,
 (\rout/R_{\sun})^{4/5}~,
\end{displaymath}
what corresponds to  $t_\mathrm{tr}\approx 34^\den$
and $t_0\approx 17^\den$ for $\alpha=0.3$ (left small arrow).
We call $t_\mathrm{tr}$ ``moment of transition''.
The transition ends at the time
$t\approx 21 ^\den (m_\mathrm{x}/3)^{2/5} \,\alpha^{-4/5}\,(\mu/0.5)^{3/5}\,
 (\rout/R_{\sun})^{4/5}$ when the
solutions match at $r=0.25\,\rout$~-- arrow B at $55^\den$ in Fig. \ref{lumin}.

This picture is reliable and useful, even though it implies
the existence of two separate
regimes, which is evidently not quite true. Indeed, at any epoch the
inner part of the disk would be scattering dominated, the
lower the accretion rate, the smaller this part.
Obtaining of an exact solution  needs consideration of
combined free-free and Thomson opacity of the gas.

There is some $t$ which corresponds to the Eddington limit
$L_\mathrm{Edd}\approx 1.3\times 10^{38} \,\mx$~erg~s$^{-1}$.
This means that the
real source evolution could be described in our model only 
at later $t$. Thus, generally speaking, the solution before this moment
appears inapplicable. As seen in Fig.~\ref{lumin},
the applicable part of the solution belongs almost entirely
to the free-free opacity regime (the bold dashed line).

The second intersection of the curves in
Fig.~\ref{lumin} at $t\approx 95^\den$ (right small arrow)
corresponds to the other intersection of functions
$F_1(\xi,t)=F_2(\xi,t+t_0)$, meanwhile the physical parameters  of the disk
calculated using formulae (\ref{d_tomson})--(\ref{tau_free})
are different. Thus the disk is at the same (free-free)
opacity regime as before.

When $\tc$ decreases to the value $\sim 10^4$~K,
the convection (which presumably appeares in the
zones of partial ionization) starts to influence the disk's
structure, and the diffusive type of radiation transfer, which we use,
is no longer valid.
For $m_\mathrm{x}=3$ and $\alpha=0.3$ this happens at $t\approx 190^\den$\,:
$t+t_0\approx 32^\mathrm{d}\,m_\mathrm{x}^{1/2}\,
\alpha^{-1}\,(\mu/0.5)\,(\rout/R_{\sun})^{1/2}\,
(r/\rout)^{-9/10}\,~ ^\mathrm{f}f^{3/10}\,~
^\mathrm{f}\Pi_3
$~.
For investigation of the disk evolution on larger time-scales
see e.g. Cannizzo et al.~(\cite{Cann_etal95}), Cannizzo~(\cite{Cann98}),
Kim et al.~(\cite{Kim_etal99}).

\section{Observed light curves}
As we mentioned in \S~6, the observed light curves can have a
slope of decline which is  {\em different} from
that of the bolometric light curves due to particular
spectral distribution. In this section we
are going to illustrate this
suggestion assuming the simplest spectral distribution of the disk emission.

To calculate the spectra, one can assume the quasi-stationary
accretion rate in the inner parts of the disk
because the $\dot M$ variation is small
there ($\dot M\propto f'(\xi)$, see Fig.~\ref{fshtrih}).
The outer parts of the disk, where
accretion rate varies significantly, contributes to the low-frequency band
of the spectrum. We discuss the X-ray band and the most luminous parts
of the disk and, thus, we take
$\dot M(t)=\dot M(0,t)$ given by (\ref{Mdotsol}).

Provided each ring in the  disk emits as a black body,
the temperature of the ring can be found as follows:
\begin{equation}
\sigmasb\, T^4 = -\,\frac{1}{2}\, W_{r\varphi}\, r\,
\frac{\mathrm{d}\,\omegak}{\mathrm{d}r} =
 \frac{3}{4}\, \omegak\, W_{r\varphi}~,
\label{temper}
\end{equation}
where
$\sigma_\mathrm{SB}=
5.67\times 10^{-5}~\mathrm{erg}~\mathrm{cm}^{-2}
\mathrm{s}^{-1}\mathrm{K}^{-4} $~ is the Stephan-Boltzmann constant.
Then
\begin{equation}
T(r,t) = \left(\frac{3\, G\,M\, \dot M(t)}{8\,\pi\, \sigmasb\,r^3}\,
             \left\{ 1-\sqrt{\frac{r_{in}}{r}} \right\}\right) ^{1/4}~.
\label{temper1}
\end{equation}
In the last expression the stationary solution for $W_{r\varphi}$
is taken.
The black-body approximation is  satisfactory
 if $\kapf \gg \kapt$. Then the
outgoing  spectrum is the sum of Planckian contributions of
each ring of the disk and has the characteristic $1/3$ slope
 for photon energies $ \ll kT_\mathrm{max}$, where
 $T_\mathrm{max}$ is the maximum effective
 temperature of the disk~(Lynden-Bell~\cite{Lynd69}).
However, if
the Thomson scattering on free electrons contributes
substantially to the opacity,
the outgoing spectrum is modified (Shakura \& Sunyaev~\cite{Shak_Suny73}).
See e.g. Ross \& Fabian~(\cite{Ross_fabi96}) for investigation of
spectral forms
of accretion disks in low-mass X-ray binaries.

The light curve is simulated by integrating at each $t$ the spectral
density 
\begin{equation}
I_\nu = \frac {4\, \pi^2 \, h_\mathsc{p}\,\nu^3}{c^2} \,
\int\limits_{r_\mathrm{in}}^{r_\mathrm{out}}\,
\frac {r\, \mathrm{d} r} {\exp\left(h_\mathsc{p}\,\nu / k\, T(r,t)\right) -1}~
\label{spectr}
\end{equation}
over the specific frequency range using (\ref{Mdotsol})  and (\ref{temper1}),
where $h_\mathsc{p}=6.626 \times 10^{-27}$~erg~s~~ is the Planck constant.
The numerical factor in (\ref{spectr})
corresponds  to the luminosity outgoing from {\em one} side of the disk.

Explaining the observed faster-than-power decay of outbursts in soft
X-ray transients, one must take into account the specificity of the
energetic band of the detector. Naturally, the observed slope of the curve
depends on  width and location of the observing interval. The narrower
this band, the more different the observed curve could look like in
comparison with the  expected bolometric light curve.   
Of course, this difference also reflects 
the spectral distribution of energy coming from the source.

\begin{figure}
\begin{center}
\centerline{\resizebox{9cm}{!}{\rotatebox{-90}{\includegraphics{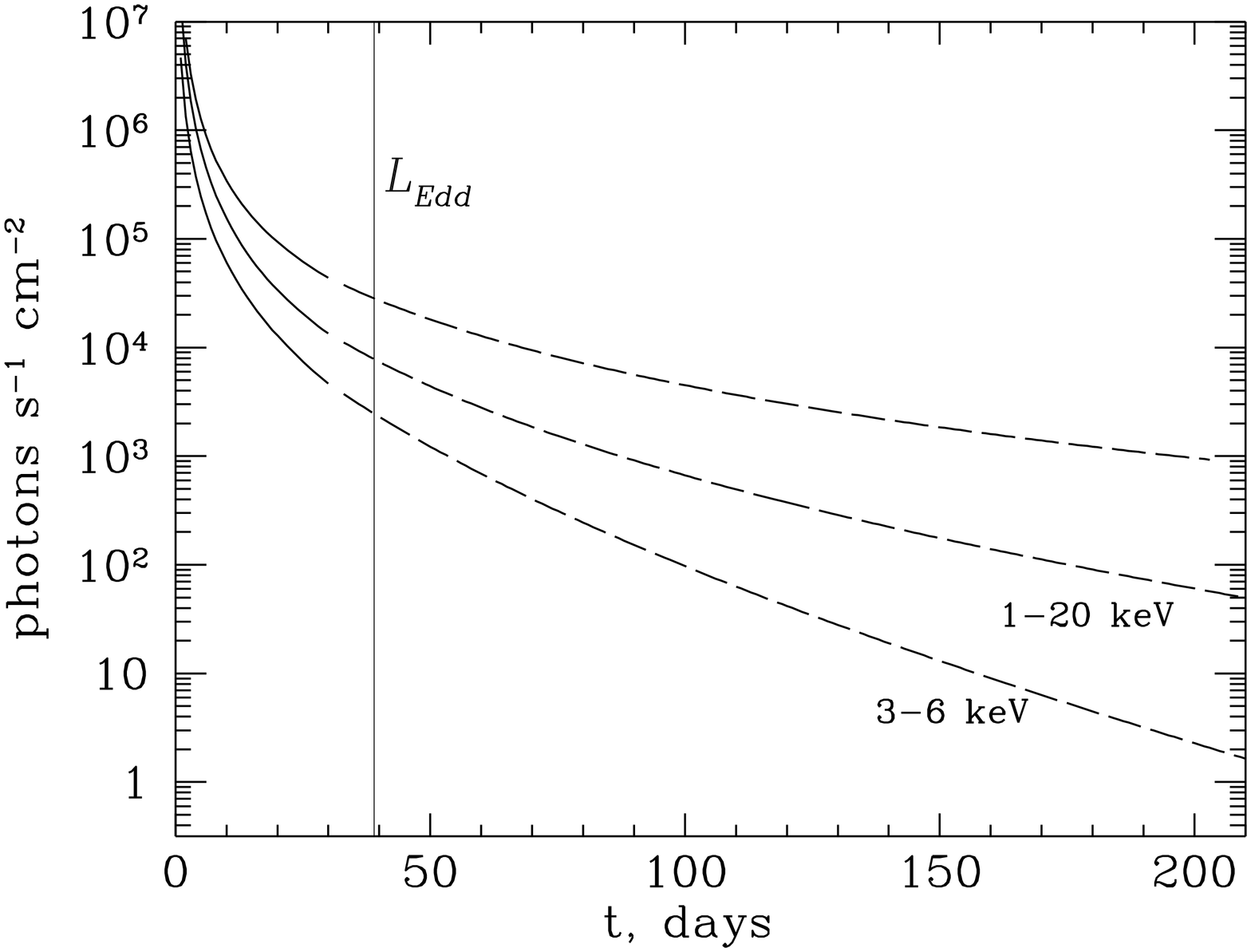}}}}
\end{center}
\vskip -0.7 cm
\caption{
The flux from one side of accretion disk at 1~kpc
for  parameters:
 $\mx=3$, $\alpha=0.3$, $\mu=0.5$, $\rout=R_{\sun}$.
The curves show the bolometric flux (upper curve), the 1--20~keV flux
 (middle curve) and  the 3--6~keV  flux (lower curve) during
 the Thomson opacity regime (solid parts) and the free-free
 opacity regime (dashed parts)
  }
 \label{flux_fig}
\end{figure}

We show here how the slope of the curve changes in the
simplest case of multi-color black body disk  spectrum according to which
spectral range is observed. Following (\ref{spectr}) we calculate
$I_\mathrm{\nu}$ and integrate it over three energy ranges:
3--6~keV, 1--20~keV and that one
in which practically all  energy is emitted.
Fig.~\ref{flux_fig} shows the photon flux variations
in two X-ray energy  ranges
(those of {\em Ariel~5} and {\em EXOSAT} or {\em Ginga} observatories) and
the bolometric flux variation for the face-on disk at an arbitrary distance
of 1~kpc.
 The vertical line marks the time
 after which bolometric luminosity of the disk's one side is
 less than $L_\mathrm{Edd}$.

One can see an almost linear trend of the X-ray flux when bolometric
luminosity is under the Eddington limit (to the right of the vertical line
in Fig.~\ref{flux_fig}),
especially in intervals of $\sim 50^\den$.
The decline becomes closer to the exponential one with time.
The slope of the curve depends on $\alpha$, $m$, $\rout$, and
other parameters.
For the same parameters as in Fig.~\ref{lumin},
the $e$-folding time falls in the range 20--30~days for the lower curve
(3--6~keV).
For instance, smaller $\alpha$ will result in less steep decline.

The natural explanation of such a result is the following: because the spectral
shape of the disk emission has Wien-form (exponential fall-off) at
the considered X-ray ranges, the law of variation of X-ray flux
is roughly proportional to $\exp\,(-h_\mathsc{p}\,\nu/k\, T^\mathrm{eff}(t))$.
In the free-free regime of opacity we have 
$T^\mathrm{eff}(t)\propto L_\mathrm{ff}^{1/4}(t)\propto \dot M(t) ^{1/4}
\propto t^{-10/12}$~(see \S~5.1). Consequently,
the observed X-ray flux varies like
 $\exp\,(-t^{5/6})$, which is quite close to exponential behavior.
We restrict ourselves to this brief and general discourse,
as  a detailed  application of our model to observed sources
is not a goal of this paper.

\section{Viscous evolution of advective disk}
As we know, the structure of an accretion disk in the vertical direction,
the relation between the viscous
tensor and the surface density in particular, defines the type
of its temporal evolution. In advective disks, which are the low-radiative
accretion flows, the relations between their characteristic
physical parameters differ significantly from those in standard disks.
In this section,
we discuss the results of \S~2 as applied to the disks
which radial structure was presented by Spruit et al.~(\cite{Spru_etal87})
and Narayan \& Yi~(\cite{Nara_yi94}, \cite{Nara_yi95}, hereafter NY).

The viscous stress and the surface density are related through the
kinematic coefficient of turbulent viscosity.
Integrating  the component of viscous stress tensor one obtains
(c.f. (\ref{stress_mal}) and (\ref{wrf}))\footnote{Formula
(\ref{wrf_adaf}) is corrected in comparison with the journal variant}:
\begin{equation}
W_{r\varphi}(r,t)
= 2 \int\limits_0^{Z_\mathrm{o}}\, w_{r\varphi}\, \mathrm{d}Z =
-2 \int\limits_0^{Z_\mathrm{o}}{\rho \nu_\mathrm{t}\,
\frac{\partial \omega}{\partial r}\,r \, \mathrm{d}Z}=
-\,\frac{\partial \omega}{\partial r}\,r \,\sigmo\,\bar{\nu}_\mathrm{t}\, ,
\label{wrf_adaf}
\end{equation}
where $\bar{\nu}_\mathrm{t}$ is the averaged kinematic coefficient of
turbulent viscosity. Then the relation between  $\sigmo(h,t)$  and
$F=W_{r\varphi}\, r^2$ is given  by
\begin{equation}
F= \left( 2\,\frac{h_\ast}{h} -
\frac 12\,\left[ \frac{\partial h_\ast}{\partial h}\right] \right)\,
h\, \sigmo\, \bar{\nu}_\mathrm{t}~.
\label{Sigma-F}
\end{equation}
Recall that $h_\ast$ is the real specific angular momentum and
$h$ is the Keplerian one.
It can be seen that $\bar{\nu}_\mathrm{t}(h,t)$ and $h_\ast(t)$ 
define what class of solutions Eq.~(\ref{basic}) will have.

If one adopts for the structure of advection-dominated accretion flow (ADAF)
the self-similar solution by NY, it can be easily inferred that
such disks exhibit the exponential with time behaviour. 
The solution of NY is given by:
\begin{equation}
 v_r = -c_1\,  \omegak\,r~,  \qquad
 \omega = c_2 \,\omegak~, \qquad
 a_\mathrm{s}^2 = c_3 \,\omegak^2 \,r^2~.
\label{adaf}
\end{equation}
Expressing $\sigmo$ in the basic Eq.~(\ref{basic})
in terms of $F$,
we obtain from (\ref{Sigma-F}) and (\ref{adaf}):
\begin{equation}
\frac{\partial F}{\partial t} = \frac{3}{4}\, \bar{\nu}_\mathrm{t}\,
\frac{(GM)^2}{h^2}\,
\frac{\partial^2 F}{\partial h^2}~.
\label{basic1}
\end{equation}
Solution (\ref{adaf}) enables deriving the relation between
$\bar{\nu}_\mathrm{t}$ and $h=\omegak\, r^2$ using the
$\alpha$ prescription of viscosity:
\begin{equation}
w_{r\varphi}=-\,\bar{\nu}_\mathrm{t}\,\rho\,r\,\frac{\de\omega}{\de r}
= \frac 32 \,\bar{\nu}_\mathrm{t}\, \rho\, c_2\,\omegak
= \alpha\, \rho\,a_\mathrm{s}^2             ~,
\label{stress_mal_adaf}
\end{equation}
where $a_\mathrm{s}$ is the isothermal sound speed. Thus $\bar{\nu}_\mathrm{t}$
is a function of radius alone,
\begin{equation}
\frac{\bar{\nu}_\mathrm{t}}{h}= \frac{2\,\alpha\, a_\mathrm{s}^2}
{3\,c_2 \omegak}\,\frac{1}{\omegak\, r^2 }
             \,  =\,\frac 23 \,\frac{c_3}{c_2}\,\alpha~,
\label{nu_t_h}
\end{equation}
and Eq.~(\ref{basic1}) can be  rewritten in the form:
\begin{equation}
\frac{\partial F}{\partial t} = \frac{D_\mathrm{a}}{h}\,
\frac{\partial^2 F}{\partial h^2}~, \qquad D_\mathrm{a} =
\frac{\alpha\,c_3}{2\,c_2}\,  (GM)^2~.
\label{basic_adaf}
\end{equation}
Solution to (\ref{basic_adaf}) is sought 
as a product of two functions $f(\xi)$ and $F(t)$, with
$\xi=h/h_\mathrm{o}$, $h_\mathrm{o}$ being some value of $h$:
 \begin{equation}
F(t) = F^\mathrm{o} \, \exp{({\lambda\,D_\mathrm{a}\,t/h_\mathrm{o}^3})}~,
\label{F(t)_adaf}
\end{equation}
\begin{equation}
\frac{d^2 f}{d \xi^2} = \lambda \,\xi\, f~.
\label{f(h)_adaf}
\end{equation}
The exponential temporal behaviour of NY flow is evident.
Generally speaking, any  disk possessing such properties of
$\bar{\nu}_\mathrm{t}$ as constancy in time would have such exponential
behaviour because its evolution would be described by a linear equation
(like  (\ref{basic_adaf})).

The question is, would the confined  NY disk keep such properties
or it would not. The fact is that NY solution describes the infinite disk.
Either the boundary conditions  destroy the linearity of
(\ref{basic_adaf})  or just the characteristic decay time changes,
 this problem requires further accurate numerical  
investigation.
For instance,
Narayan et al.~(\cite{Nara_etal97}) calculated numerically the global structure
of stationary
advection-dominated flow with consistent boundary conditions; they
noted that although the self-similar solution (\ref{adaf})
makes significant errors close to the boundaries, it gives
the reasonable description of the overall properties of the flow.

Further we assume that exponential trend of solution persists.
Generally speaking, the equation determining $f(\xi)$ will differ
from (\ref{f(h)_adaf}). This difference may be not very significant.
One can see that Eq.~(\ref{f(h)_adaf}) is a particular case
of (\ref{f(h)SS}) where  $n=1$ and $m=0$ and, hence, the solution can
be found according to (\ref{SolSS}) and (\ref{f(xi)}).
Besides, the solution of (\ref{f(h)_adaf})
can be found in terms of Airy functions
(Bessel functions of  order $1/3$).

The accretion rate
evolves with time as follows (c.f. (\ref{Mdotsol}) and (\ref{F(t)_adaf})):
\begin{equation}
\dot M \propto \exp{(\lambda\,D_\mathrm{a}\,t/h_\mathrm{o}^3 )}~.
\label{accr_adaf}
\end{equation}
The value of accretion rate can be determined if an initial condition
is imposed at some $t$.
Mahadevan~(\cite{Maha97}) showed that ADAF luminosity $\propto \dot M^2$
or $\propto \dot M$
according to whether the electron heating is dominated by
the Coulomb interactions or by the viscous friction.
Subsequently, the luminosity has an exponential decay too.

We can estimate the characteristic time of evolution of such flow.
It can be obtained from (\ref{accr_adaf}). Let us compare the diffusion time
$t_\mathrm{ev}\simeq r^2/\bar{\nu}_\mathrm{t}$ with the corresponding orbital
period $2\pi/\omega$.
Since $\bar{\nu}_\mathrm{t} \simeq \alpha \,a_\mathrm{s}^2/\omega$,
with (\ref{adaf}) we have:
\begin{equation}
\frac{t_\mathrm{ev}}{t_\mathrm{orb}} \,
= \frac{1}{\alpha}\, \frac{c_2^2}{2\,\pi\,c_3}
= \frac{1}{2\,\pi\,\alpha}\, \frac{5/3 - \gamma}{(\gamma-1)\,f_\mathrm{a}}
\,.
\label{vremya_adaf}
\end{equation}
We use the expressions for $c_2$ and $c_3$ from NY.
Here $\gamma$  is the ratio of specific heats; $f_\mathrm{a}$ measures
the efficiency of radiative cooling. In the limit of no radiative cooling,
we have $f_\mathrm{a}=1$ while in the opposite limit of very efficient
cooling $f_\mathrm{a}=0$. NY solution is degenerate
if $\gamma=5/3$ because the angular velocity of the flow is zero in this case.

We can see that the time-dependent advection-dominated disk is quickly
depleted  if $\alpha$ is not small. For example, consider
the light curves of \hbox{X-ray}
novae which have the exponential decay time scales 
$\sim  30^{\mathrm{d}}$.
To obtain $t_\mathrm{ev}$ of such order, $\alpha$ should be $\sim 10^{-2}$.
However, the advection-dominated solution  ceases to exist if the accretion
rate is greater than the critical value
$\dot M_\mathrm{crit}\sim \alpha^2\,\dot M_\mathrm{Edd} $~
(Narayan \& Yi~\cite{Nara_yi95}, Mahadevan~\cite{Maha97}), where
$\dot M_\mathrm{Edd}= L_\mathrm{Edd}\,(\eta/0.1)^{-1}\,c^{-2}\,=
1.39\times 10^{18}\, m_\mathrm{x}$.
Hence, $\alpha\sim 10^{-2}$ yields the critical accretion rate
$\dot M_\mathrm{crit}\sim 10^{14}\, m_\mathrm{x}$~g~s$^{-1}
\sim 10^{-4} \dot M_\mathrm{Edd}$.

\section{Discussion and conclusion}

In this work, we presented the analytical solutions to time-dependent
accretion in binary systems. During an outburst we propose the specific
external boundary conditions on a disk confined due to tidal interactions.
For two opacity regimes the full analytical time-dependent solutions for
the Keplerian disk are obtained and an
{\em asymptotic} light curve is calculated
with smooth transition between opacity regimes.
During the decline phase accretion disks around black holes appear to be
dominated by the free-free and free-bound opacity
in order to comply with the Eddington limit on luminosity. This phase
is characterized by the power-law decay of accretion rate
$\propto t^{-10/3}$. It is shown that the decay time scale depends 
on the real energetic band of detector (Fig.~\ref{flux_fig}).

The results obtained in this work
can be  applied  to the accreting systems having variable emission of
flare type if emission is essentially due to the fully ionized accretion
disk around a
black  hole, or a neutron star, or a white dwarf. 

Narayan \& Yi~(\cite{Nara_yi94}) accretion flows are shown to
undergo exponential decays if the disk has infinite  size.
This notable result probably persists even when the advective disk is
in a binary system. The latter suggestion is to be thoroughly
considered in the accurate numerical investigation. If this is the case,
the abrupt steep falls
observed in several \hbox{X-ray} novae (Tanaka \&
Shibazaki~\cite{Tana_shib96}) in the last phase of the decay,
at luminosity levels $\lesssim 10^{36}$~erg s$^{-1}$,
can be interpreted in terms of quickly depleting ADAF (\S~8) with
relevant values of $\alpha \sim 10^{-1}$.

Using the results of this work, we can explain the general features
of \hbox{X-ray} novae light curves in the early phase.
Typical XN outburst light
curves (see Tanaka \& Shibazaki~\cite{Tana_shib96};
Chen at al.~\cite{Chen_etal97} for a
review) show quasi-exponential decay. To date several
approaches have been used to account for XN features.
The exponential decays were obtained in the framework
of disk instability model (Cannizzo et al.~\cite{Cann_etal95};
Vishniac~\cite{Vish97}; Cannizzo~\cite{Cann98}) in which
the large time-scale evolution of the disk is considered.
Mineshige et al.~(\cite{Mine_etal93}) argued that the exponential decays in XN
can be reproduced if the mass and the angular momentum are
efficiently removed from the inner portions of the disk at a constant
rate, or wind mass loss or enhanced tidal dissipation
could be substantial.
King \& Ritter~(\cite{King_ritt98}) took into account the irradiation of the disk
by the central \hbox{X-ray} source and obtained the characteristic XN light curves.  

We suggest an alternative reason to explain this remarkable feature,
at least during the early stages of the outburst when the disk
is fully ionized.
Nearly exponential \hbox{X-ray} decays $\propto\exp\,(-t^{5/6})$ are obtained
taking into account the fact that the \hbox{X-ray} light curves are observed in 
the energetic range where the spectrum of the disk has Wien-form.
Black hole XN spectra typically are composed of an ultrasoft component
and a hard power-law component (e.g. Tanaka~\cite{Tana92};
Tanaka \& Shibazaki~\cite{Tana_shib96}).
At the first stages after outburst the ultrasoft component dominates
and can be represented by a multicolor blackbody disk (Tanaka~\cite{Tana92}).
This component has an exponential fall-off, a decisive factor to
produce observed exponential trends.
The observed characteristic
times can be obtained within reasonable intervals of parameters
(Fig.~\ref{flux_fig}, \S~7).

The secondary peak commonly observed in XN can be qualitatively
analytically produced by 
certain reconstruction of viscosity mechanisms and corresponding increase
of $\alpha$  ({\S~6.1}).
Possible mechanisms of reflares involving irradiation effects
were investigated by Kim et al.~(\cite{Kim_etal94}), Mineshige~(\cite{Mine94}),
King \& Ritter~(\cite{King_ritt98})
(see, however, Cannizzo~\cite{Cann98}).

Of course, the accretion disk spectrum represents only one contribution
to the total observed spectrum of the source. The corona around the    
disk is probably responsible for the other spectral components.        
In addition, taking into consideration the irradiation of the outer
parts of the disk would affect evolution of the disk (see, e.g.
King \& Ritter~\cite{King_ritt98}; Kim et al.~\cite{Kim_etal99}).
Kim et al.~(\cite{Kim_etal99}) constructed
an optical light curve of a XN and found the
direct irradiation of the disk by the inner layers to have only a small
effect on the outer disk because of shadowing. The indirect
irradiation (from a corona or a chromosphere above the disk) is found
to affect the light curve more strongly. We suggest that the irradiation
of the twisted warped disk could also 
result in important heating of the outer layers.
Further investigation and applications to observed sources will
be the basis of our future work.

\begin{acknowledgements}
We are grateful to the anonymous referee for helpful arguments and comments.
 This work is partially  supported by the RFBR grant \hbox{98--02--16801} and
the program `Universitety Rossii' (grant 5559) of the Ministry of Teaching
and Professional Education, Russia. GVL is thankful to RFBR project 
`Molodye uchenye Rossii' of 1999.
\end{acknowledgements}


\begin{thebibliography}{}

\bibitem[1995]{Cann_etal95}
Cannizzo, J.K., Chen, W., Livio, M., 1995, ApJ 454, 880

\bibitem[1998]{Cann98}
Cannizzo, J.K., 1998, ApJ 494, 366

\bibitem[1997]{Chen_etal97}
Chen, W., Shrader, C.R., Livio, M., 1997, ApJ 491, 312

\bibitem[1984]{Fili84}
Filipov, L.G., 1984, Advances in Space Research 3, 305

\bibitem[1994]{Ichi_osak94}
Ichikawa, S., Osaki, Y., 1994, PASJ 46, 621

\bibitem[1998]{Kato_etal98}
Kato, S., Fukue, J., Mineshige, S., 1998, Black-hole Accretion
Disks. Kyoto University Press, Japan

\bibitem[1998]{Kets_shak98}
Ketsaris, N.A., Shakura,  N.I., 1998, Astronomical and Astrophysical
Transactions 15, 193

\bibitem[1994]{Kim_etal94}
Kim, S.-W., Wheeler, J.C., Mineshige, S., 1994, American Astr. Soc. 185, 1109

\bibitem[1999]{Kim_etal99}
Kim, S.-W., Wheeler, J.C., Mineshige, S., 1999, PASJ 51, 1999

\bibitem[1998]{King_ritt98}
King, A.R., Ritter, H., 1998, MNRAS 293, L42

\bibitem[1952]{Lust52}
L\"ust, R., 1952, Naturforsch 7a, 87

\bibitem[1969]{Lynd69}
Lynden-Bell, D., 1969, Nature 233,  690

\bibitem[1974]{Lynd_prin74}
Lynden-Bell, D., Pringle, J.E., 1974, MNRAS 168, 603

\bibitem[1987]{Lyub_shak87}
Lyubarskii, Yu.E., Shakura, N.I., 1987, SvA 13, 386

\bibitem[1982]{Meye_meye82}
Meyer, F., Meyer-Hofmeister, E., 1982, A\&A 106, 34

\bibitem[1997]{Maha97}
Mahadevan, R., 1997, ApJ 477, 585

\bibitem[1978]{Miha78}
Mihalas, D., 1978, In: Freeman W.H. (ed.) Stellar Atmospheres,
San Francisco

\bibitem[1993]{Mine_etal93}
Mineshige, S., Yamasaki, T., Ishizaka, C., 1993, PASJ 45, 707

\bibitem[1994]{Mine94}
Mineshige, S., 1994, ApJ 431, L99

\bibitem[1995]{Naka_kato95}
Nakao, Y., Kato, S., 1995, PASJ 47, 451

\bibitem[1994]{Nara_yi94}
Narayan, R., Yi, I., 1994, ApJ 428, L13

\bibitem[1995]{Nara_yi95}
Narayan, R., Yi, I., 1995, ApJ 452, 710

\bibitem[1997]{Nara_etal97}
Narayan, R., Kato, S., Honma, F., 1997, ApJ 476, 49

\bibitem[1999]{Ogil99}
Ogilvie, G.I., 1999, MNRAS 306, L90

\bibitem[1977]{Pacz77}
Paczy\'nski, B., 1977, ApJ 216, 822

\bibitem[1977]{Papa_prin77}
Papaloizou, J., Pringle, J.E.,  1977, MNRAS 181, 441

\bibitem[1996]{Ross_fabi96}
Ross, R.R., Fabian, A.C., 1996, MNRAS 281, 637

\bibitem[1972]{Shak72}
Shakura, N.I., 1972, AZh 49, 921

\bibitem[1973]{Shak_Suny73}
Shakura, N.I., Sunyaev, R.A., 1973, A\&A 24,  337

\bibitem[1987]{Spru_etal87}
Spruit, H.C., Matsuda, T., Inoue, M., Sawada, K., 1987, MNRAS 229, 517

\bibitem[1992]{Tana92}
Tanaka, Y., 1992, In: Makino F., Nagase F. (eds.)
Ginga Memorial Symp., ISAS, 19

\bibitem[1996]{Tana_shib96}
Tanaka, Y., Shibazaki, N., 1996, ARA\&A 34, 607 

\bibitem[1980]{Tayl80}
Tayler, R.J.,  1980, MNRAS 191, 135

\bibitem[1997]{Vish97}
Vishniac, E.T., 1997, ApJ 482, 414

\bibitem[1948]{Weiz48}
Weizs\"acker, C.F., 1948, Z. Naturforsch 3a, 524

\bibitem[1996]{Zayc_poly96}
Zaycev, V.F., Polyanin, A.D., 1996, Handbook of Partial Differential Equations.
Moscow

\bibitem[1967]{Zeld_raiz67}
Zeldovich, Ya.B., Raizer, Yu.P., 1967,
Physics of Shock Waves and High-Temperature Hydrodynamic
Phenomena. New York: Academic Press

\bibitem[1969]{Zeld_shak69}
Zeldovich, Ya.B., Shakura, N.I., 1969, AZh 46, 225
\end{thebibliography}
\end{document}